\documentclass[letterpaper, 10 pt, conference]{ieeeconf}  

\IEEEoverridecommandlockouts                              

\usepackage{url}
\usepackage{balance}
\usepackage[normalem]{ulem}
\useunder{\uline}{\ul}{}
\usepackage{supertabular,booktabs}
\usepackage{tabularx}
\usepackage{longtable}
\usepackage{graphicx}
\usepackage{xcolor}
\usepackage{adjustbox}
\usepackage{multirow}
\usepackage{makecell}
\usepackage{longtable}
\usepackage{supertabular}
\usepackage{tabularx, booktabs, longtable}
\usepackage[labelformat=simple]{subcaption}

\usepackage{etoolbox}
\AtBeginEnvironment{tabular}{\tiny}
\newcolumntype{C}{>{\centering\arraybackslash}X} 
\usepackage[acronym]{glossaries}
\newacronym{mtl}{MTL}{Multi-task Learning}
\newacronym{gan}{GAN}{Generative Adversarial Network}
\newacronym{fcn}{FCN}{Fully Convolutional Network}
\newacronym{gam}{GAM}{Gate Attention Module}
\newacronym{dam}{DAM}{Decoder Attention Module}
\newacronym{rab}{RAB}{Residual Attention Block}
\newacronym{ra}{RA}{Reverse Attention}
\newacronym{idc}{IDC}{Improved Dilation Convolution}
\newacronym{ag}{AG}{Attention Gate}
\newacronym{tasegnet}{TA-SegNet}{Tri-level Attention-based Segmentation Network}
\newacronym{tau}{TAU}{Tri-level Attention Unite}
\newacronym{qapnet}{QAP-Net}{Quadruple Augmented Pyramid Network}
\newacronym{raiunet}{RAIU-Net}{Residual Attention Inception U-Net}
\newacronym{chsnet}{CHS-Net}{Covid-19 Hierarchical Segmentation Network}
\newacronym{csse}{CSSE}{Systems Science and Engineering}
\newacronym{jhu}{JHU}{Johns Hopkins University}
\newacronym{rtpcr}{RT-PCR}{Reverse-Transcription Polymerase Chain Reaction}
\newacronym{ct}{CT}{Computed Tomography}
\newacronym{fpn}{FPN}{Feature Pyramid Network}
\newacronym{pspnet}{PSPNet}{Pyramid Scene Parsing Network}
\newacronym{manet}{MA-Net}{Multi-scale Attention Net}
\newacronym{se}{SE-Net}{Squeeze-and-Excitation Network}
\newacronym{ggo}{GGO}{Ground Glass Opacity}
\newacronym{iou}{IoU}{Intersection over Union}

\usepackage{tikz}
\usepackage{textcomp}
\usepackage{hyperref}
\usepackage{lipsum}

\newcommand\copyrighttext{%
  \footnotesize \textcopyright 2021 IEEE. Personal use of this material is permitted.
  Permission from IEEE must be obtained for all other uses, in any current or future
  media, including reprinting/republishing this material for advertising or promotional
  purposes, creating new collective works, for resale or redistribution to servers or
  lists, or reuse of any copyrighted component of this work in other works.
  }
\newcommand\copyrightnotice{%
\begin{tikzpicture}[remember picture,overlay]
\node[anchor=south,yshift=10pt] at (current page.south) {\fbox{\parbox{\dimexpr\textwidth-\fboxsep-\fboxrule\relax}{\copyrighttext}}};
\end{tikzpicture}%
}

\title{\LARGE \bf
Spark in the Dark: Evaluating Encoder-Decoder Pairs for COVID-19 CT's Semantic Segmentation 
}

\author{
        Bruno A. Krinski, Daniel V. Ruiz, and Eduardo Todt \\
        Department of Informatics, Federal University of Paran\'a (UFPR), Curitiba, PR, Brazil \\ 
         \textit{\{bakrinski, dvruiz, todt\}@inf.ufpr.br}
        } 

\begin{document}
\maketitle
\copyrightnotice
\thispagestyle{empty}
\pagestyle{empty}

\begin{abstract}

With the COVID-19 global pandemic, computer-assisted diagnoses of medical images have gained a lot of attention, and robust methods of Semantic Segmentation of \gls*{ct} turned highly desirable. Semantic Segmentation of \gls*{ct} is one of many research fields of automatic detection of Covid-19 and was widely explored since the Covid-19 outbreak. In the robotic field, Semantic Segmentation of organs and \glspl*{ct} are widely used in robots developed for surgery tasks. As new methods and new datasets are proposed quickly, it becomes apparent the necessity of providing an extensive evaluation of those methods. To provide a standardized comparison of different architectures across multiple recently proposed datasets, we propose in this paper an extensive benchmark of multiple encoders and decoders with a total of 120 architectures evaluated in five datasets, with each dataset being validated through a five-fold cross-validation strategy, totaling 3.000 experiments. To the best of our knowledge, this is the largest evaluation in number of encoders, decoders, and datasets proposed in the field of Covid-19 \gls*{ct} segmentation.

\end{abstract}

\section{INTRODUCTION}

As of late 2019, the world faces the worst pandemic in years, with the new coronavirus disease, COVID-19, becoming a threat worldwide~\cite{Wang2020}. According to the global case count from the \gls*{csse} at \gls*{jhu} (updated April 20th, 2021), there are a total of 196,743,788 cases identified all around the globe, with a total of 4,201,812 global deaths~\cite{coviddata}. 


Early diagnosis is one of the most effective ways to fight against the virus~\cite{Chen2020.04.06.20054890},  with automatic detection of Covid-19 presence in \gls*{ct} being highly desirable, and recent results showing effectiveness in diagnosing and identifying Covid-19 patients~\cite{Shi2021}. Semantic Segmentation~\cite{Cao2020} of \gls*{ct} is one of many research fields of automatic detection of Covid-19 and was widely explored since the Covid-19 outbreak~\cite{Shi2021}. In the robotic field, Semantic Segmentation of organs and \glspl*{ct} are widely used in robots developed for surgery tasks~\cite{Muradore2015, Li2021}.

Aiming to perform the segmentation of Covid-19 \glspl*{ct}, many studies apply Deep Learning techniques and Deep Neural Networks, achieving impressive results in the task~\cite{Shi2021}. Deep Neural Networks are widely applied in segmentation problems due to their great generalization capacity, learning to represent different classes of objects~\cite{GarciaGarcia2018, krinski2019}. 

However, with new approaches being proposed quickly, an urgency aggravated by the global pandemic, the need for an extensive evaluation becomes apparent. To provide a standardized comparison of different architectures across multiple recently proposed datasets, we propose in this paper an extensive benchmark of multiple encoders and decoders with a total of 120 architectures evaluated. The models were trained and evaluated across five different \glspl*{ct} datasets: MedSeg~\cite{medseg}, Zenodo~\cite{zenodo}, CC-CCII~\cite{Zhang2020}, MosMed~\cite{Morozov2020}, Ricord1a~\cite{ricord1a}. Each dataset was validated through a five-fold cross-validation strategy, totaling 3.000 experiments. To the best of our knowledge, this is the largest evaluation in number of encoders, decoders, and datasets proposed in the field of Covid-19 \gls*{ct} segmentation.

All models were trained using the same loss function to ensure baseline comparability, see section~\ref{architectures}, and without any data augmentation. The goal is to provide a lower bound estimate for each encoder-decoder combination on different datasets. A specialized study on the impact of different data augmentation techniques on each model was left for future works.




\section{Related Work}

\begin{figure*}[!ht]
\centering
\captionsetup[subfigure]{width=0.90\linewidth}
\subfloat{
	    \includegraphics[width=0.32\textwidth]{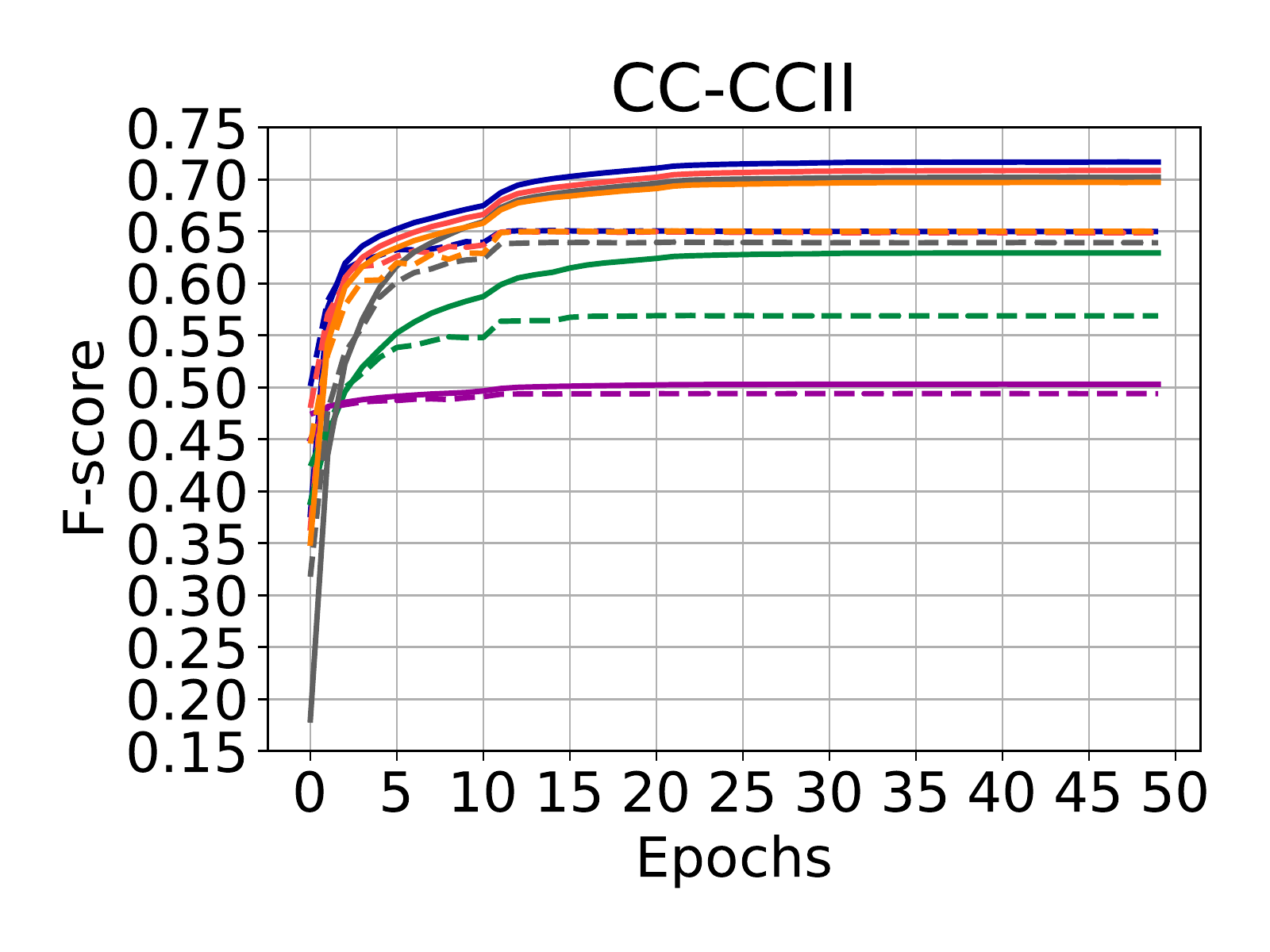}
	    \label{fig:covid19china}
	}
\subfloat{
	    \includegraphics[width=0.32\textwidth]{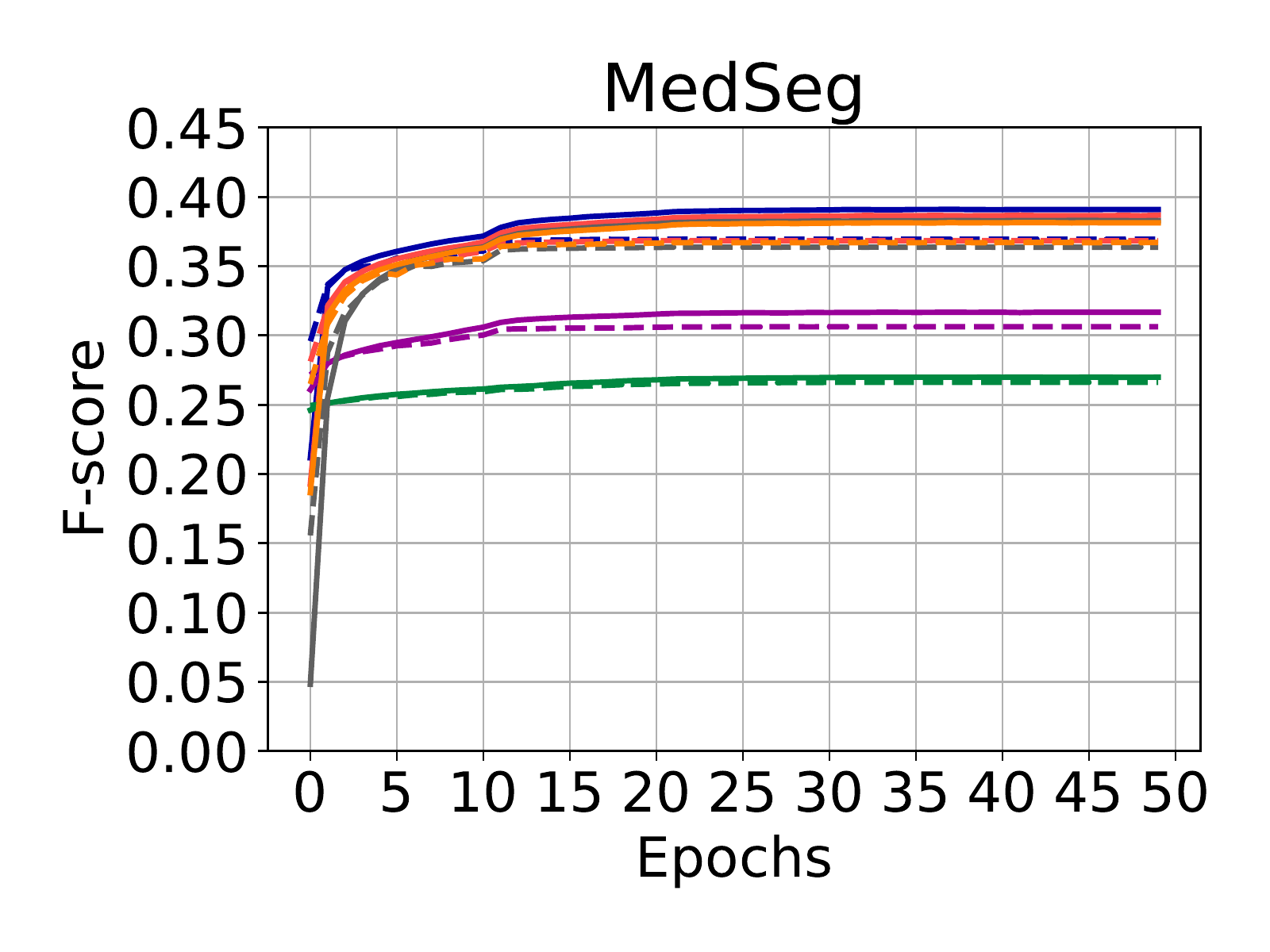}
	    \label{fig:medseg}
	}
	
\subfloat{
	    \includegraphics[width=0.32\textwidth]{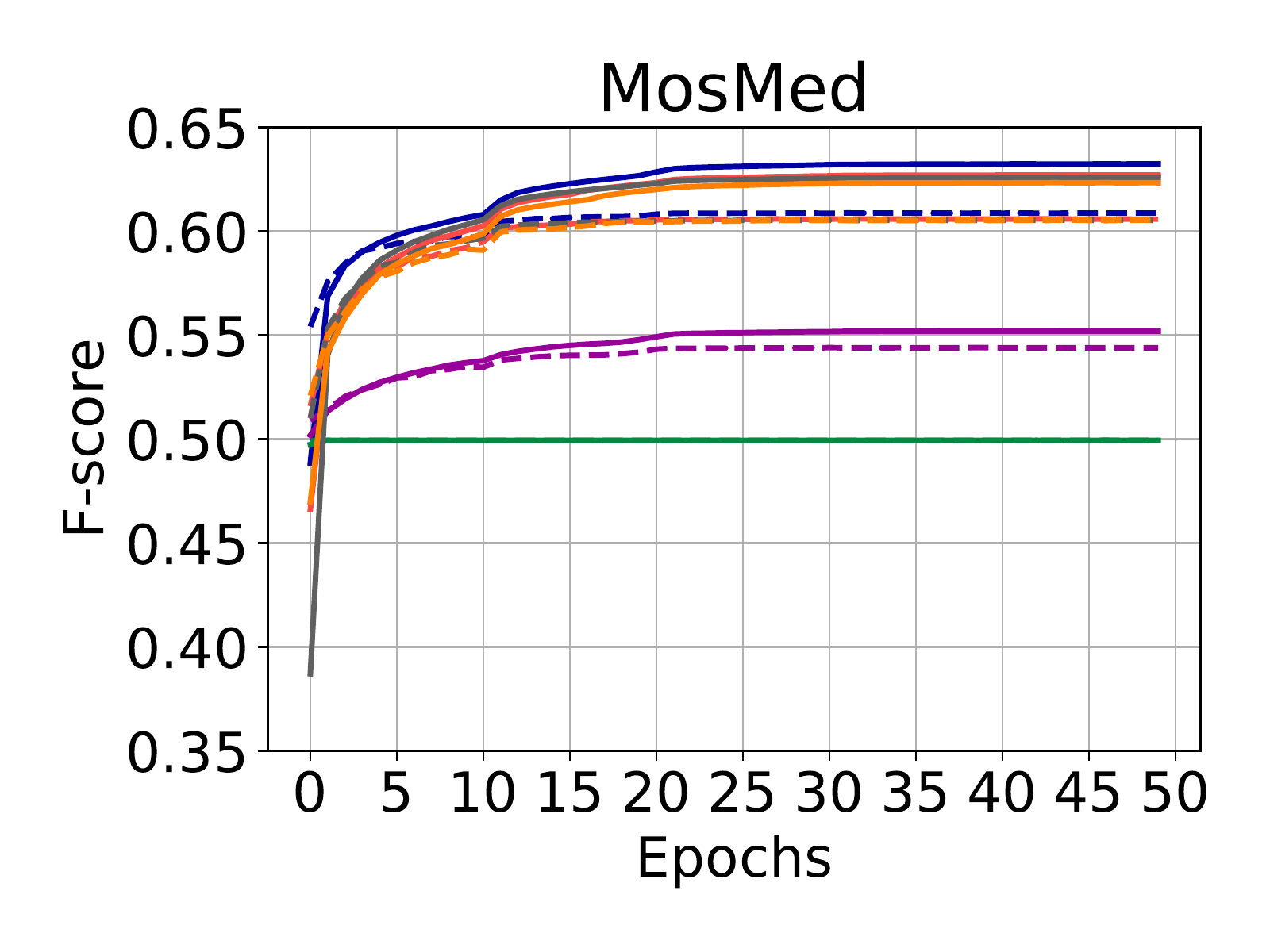}
	    \label{fig:mosmed}
	}
\subfloat{
	    \includegraphics[width=0.32\textwidth]{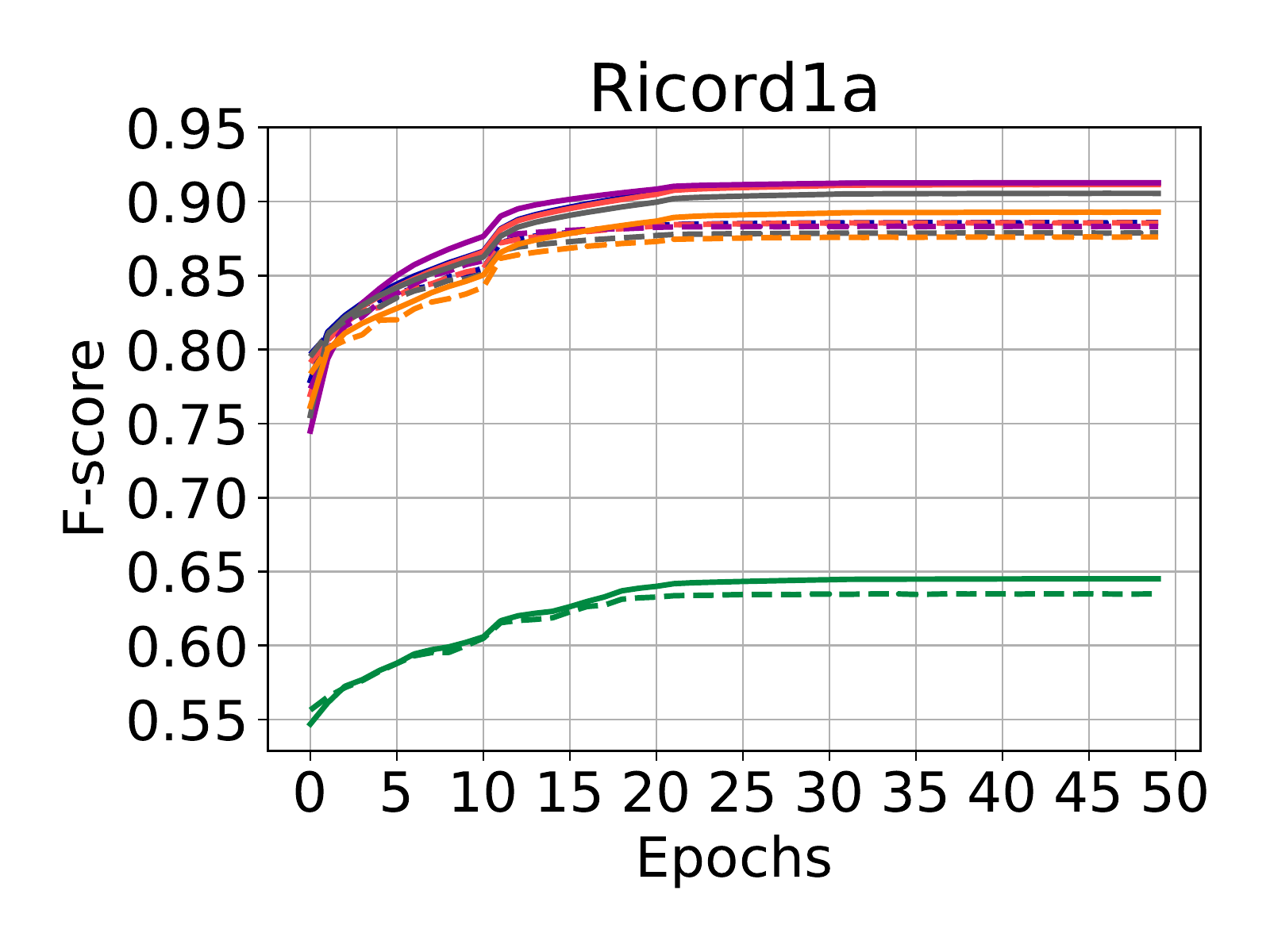}
	    \label{fig:ricord1a}
	}
\subfloat{
	    \includegraphics[width=0.32\textwidth]{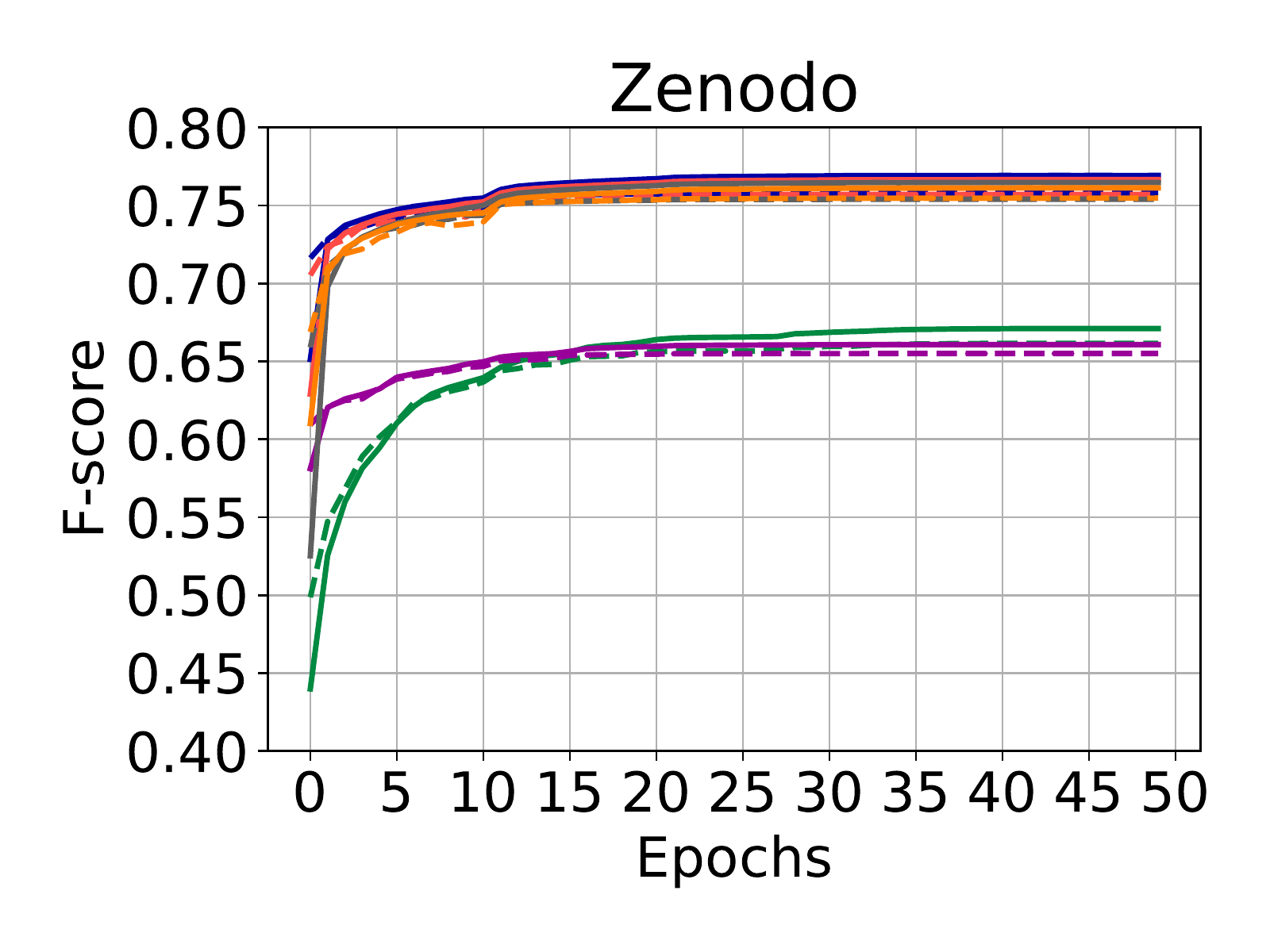}
	    \label{fig:zenodo}
	}
	
\subfloat{
	    \includegraphics[width=0.8\textwidth]{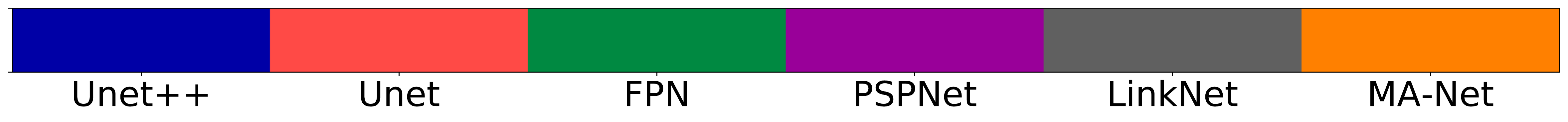}
	    \label{fig:scale}
	}
\caption[]{The train and validation average of the k-folds F-score curves for each decoder. The filled line is the train F-score curve, and the dashed line is the validation F-score curve. Each encoder-decoder combination was trained through a five-fold cross-validation strategy. The curves are color-coded by decoder (U-Net, U-Net++, \gls*{fpn}, \gls*{pspnet}, LinkNet, and \glsfirst*{manet}) and represent the average F-score of the five-folds of all encoders (ResNet-50-fold0, ..., ResNet-50-fold4, ResNet-101-fold0, ..., ResNet-101-fold4, etc.) using that decoder.}
\label{fig:training}
\end{figure*}

This section presents studies about Semantic Segmentation techniques proposed to perform segmentation of Covid-19 CT-Scans. Starting with the study presented in~\cite{Saood2021}, the authors performed a comparison between U-net~\cite{Ronneberger2015} and SegNet~\cite{Badrinarayanan2017} architectures on the Covid-19 CT-Scan segmentation problem. In~\cite{Fan2020}, the authors proposed a segmentation network with \glspl*{ra} modules attached in the decoder side. These \gls*{ra} modules aim to focus on the opposite regions of the interest regions, focusing the attention on background regions instead of Covid lesion regions.


In~\cite{zhao2021d2a}, the authors modified the U-net model by replacing the decoder side's convolution layers with dilated convolutions. Also, proposed a dual attention module attached to the skip connections in the decoder module. The dual attention module comprises a \gls*{gam} and a \gls*{dam}. The former refine the features extracted by the encoder, and the latter reduces the semantic gap by fusing high and low-level feature maps. A module called \gls*{rab} is attached to each block of the decoder side. In this \gls*{rab}, a hybrid dilated convolution module and a \gls*{dam} are integrated to refine post-upsample features. 

In~\cite{Zhou2020}, the authors proposed a U-net-like architecture with attention modules to perform segmentation of Covid-19 lesions. Each block in the encoder and decoder has a Res\_dil block after a sequence of convolutional layers. These Res\_dil blocks use dilated convolutions to help the network to learn more detailed regions of the image. Each skip connection of the U-net, which connects a convolution block to its mirror deconvolution block, has two attention mechanisms built with dilated convolutions. The first one attached at the beginning of the skip connection and another at the end. 

The architecture proposed in~\cite{JosephRaj2021} is a U-net-like model with an encoder-decoder structure. The authors proposed an \gls*{idc} module attached between the encoder and decoder to increase the receptive field and gather detailed edge information, which helps extract characteristics. An \gls*{ag} module is used in each skip connection between encoder and decoder mirror blocks to reduce the loss of spatial information in the feature mapping at the end of the encoder.



In~\cite{zhang2020exploiting}, the authors evaluated the combination of features extracted from CT-Scans with Covid lesions and features extracted from CT-Scans with lesions from no-Covid samples. In~\cite{chen2020residual}, a study evaluating U-Net~\cite{Ronneberger2015} application to the Covid CT-Scan segmentation problem is presented. All convolutional blocks of the evaluated U-net were replaced by a ResNeXt~\cite{Xie2017} block. Also, an attention mechanism was proposed to capture complex features from the feature maps. The output of each block in the decoder is concatenated with the output of an encoder block, inputs the attention mechanism, and then inputs the following decoder block. A comparative study of networks for Covid-19 problem is presented in~\cite{bizopoulos2021comprehensive}. The authors evaluated 100 architectures in two Covid-19 \gls*{ct}-Scan datasets.


\begin{table*}[!ht]
\centering
\caption{The average value of five-folds cross-validation strategy for test set evaluated with the last train weight. The best F-score values are highlighted in blue and the best IoU values are highlighted in red.} 
\resizebox{0.9\textwidth}{!}{%
\begin{tabular}{cccccccccccc}

\toprule
\multicolumn{1}{c}{\textbf{Decoder}} &
\multicolumn{1}{c}{\textbf{Encoder}} &
\multicolumn{2}{c}{\textbf{CC-CCII}} &
\multicolumn{2}{c}{\textbf{MedSeg}} &
\multicolumn{2}{c}{\textbf{MosMed}} &
\multicolumn{2}{c}{\textbf{Ricord1a}} &
\multicolumn{2}{c}{\textbf{Zenodo}} 
\\ \midrule

\makecell{} &
\makecell{} &
\makecell{F-score} &
\makecell{IoU} & 
\makecell{F-score} &
\makecell{IoU} & 
\makecell{F-score} &
\makecell{IoU} & 
\makecell{F-score} &
\makecell{IoU} & 
\makecell{F-score} &
\makecell{IoU}
\\ \midrule

\makecell{U-Net++} &
\makecell{ResNet-50\\
          ResNet-101\\
          ResNeXt50\_32x4d\\
          ResNeXt101\_32x8d\\
          Res2Net-50\_26w\_4s\\
          Res2Net-101\_26w\_4s\\
          Vgg16\\
          Densenet-121\\
          Densenet-169\\
          Densenet-201\\
          SE-ResNet-50\\
          SE-ResNet-101\\
          SE-ResNeXt50\_32x4d\\
          SE-ResNeXt101\_32x4d\\
          RegNetx-002\\
          RegNetx-004\\
          RegNetx-006\\
          RegNety-002\\
          RegNety-004\\
          RegNety-006}&
          
\makecell{0.6460\\0.6440\\0.6462\\0.6477\\0.6467\\0.6458\\0.6482\\\textbf{\textcolor{blue}{0.6516}}\\0.6505\\0.6474\\0.6482\\0.6457\\0.6479\\0.6440\\0.6414\\0.6384\\0.6386\\0.6402\\0.6402\\0.6400}&
\makecell{0.6068\\0.6043\\0.6065\\0.6078\\0.6076\\0.6061\\0.6082\\\textbf{\textcolor{red}{0.6121}}\\0.6108\\0.6074\\0.6076\\0.6057\\0.6076\\0.6042\\0.6008\\0.5981\\0.5991\\0.6001\\0.6007\\0.6001}&
\makecell{0.3732\\0.3719\\0.3740\\0.3721\\0.3744\\0.3733\\0.3717\\0.3756\\0.3750\\0.3743\\0.3767\\\textbf{\textcolor{blue}{0.3770}}\\0.3760\\0.3763\\0.3744\\0.3727\\0.3742\\0.3743\\0.3710\\0.3735}&
\makecell{0.3475\\0.3464\\0.3479\\0.3464\\0.3485\\0.3474\\0.3459\\0.3494\\0.3490\\0.3480\\0.3510\\\textbf{\textcolor{red}{0.3512}}\\0.3505\\0.3509\\0.3486\\0.3468\\0.3484\\0.3486\\0.3455\\0.3476}&
\makecell{0.6135\\0.6114\\0.6115\\0.6103\\0.6115\\0.6132\\0.6160\\0.6117\\0.6123\\0.6124\\\textbf{\textcolor{blue}{0.6219}}\\0.6205\\0.6168\\0.6192\\0.6112\\0.6127\\0.6108\\0.6134\\0.6162\\0.6146}&
\makecell{0.5880\\0.5859\\0.5861\\0.5847\\0.5863\\0.5876\\0.5913\\0.5857\\0.5862\\0.5868\\\textbf{\textcolor{red}{0.5969}}\\0.5955\\0.5927\\0.5947\\0.5862\\0.5878\\0.5860\\0.5885\\0.5907\\0.5895}&
\makecell{0.8802\\0.8697\\0.8789\\0.8721\\0.8776\\0.8790\\0.8686\\0.8873\\0.8836\\0.8814\\0.8908\\0.8910\\0.8963\\0.8963\\0.8954\\0.8909\\0.8940\\0.8954\\\textbf{\textcolor{blue}{0.8983}}\\0.8962}&
\makecell{0.8185\\0.8061\\0.8172\\0.8092\\0.8158\\0.8176\\0.8053\\0.8269\\0.8228\\0.8207\\0.8308\\0.8311\\0.8377\\0.8378\\0.8369\\0.8317\\0.8354\\0.8368\\\textbf{\textcolor{red}{0.8407}}\\0.8378}&
\makecell{0.7351\\0.7338\\0.7352\\0.7344\\0.7353\\0.7350\\0.7357\\0.7358\\0.7354\\0.7359\\0.7360\\0.7358\\\textbf{\textcolor{blue}{0.7367}}\\0.7366\\0.7360\\0.7364\\0.7352\\0.7364\\0.7364\\0.7358}&
\makecell{0.7065\\0.7052\\0.7069\\0.7058\\0.7071\\0.7066\\0.7075\\0.7077\\0.7070\\0.7078\\0.7078\\0.7077\\\textbf{\textcolor{red}{0.7091}}\\0.7088\\0.7079\\0.7085\\0.7069\\0.7084\\0.7086\\0.7076}
          
\\ \midrule

\makecell{U-Net} &
\makecell{ResNet-50\\
          ResNet-101\\
          ResNeXt50\_32x4d\\
          ResNeXt101\_32x8d\\
          Res2Net-50\_26w\_4s\\
          Res2Net-101\_26w\_4s\\
          Vgg16\\
          Densenet-121\\
          Densenet-169\\
          Densenet-201\\
          SE-ResNet-50\\
          SE-ResNet-101\\
          SE-ResNeXt50\_32x4d\\
          SE-ResNeXt101\_32x4d\\
          RegNetx-002\\
          RegNetx-004\\
          RegNetx-006\\
          RegNety-002\\
          RegNety-004\\
          RegNety-006}&
          
\makecell{\textbf{\textcolor{blue}{0.6512}}\\0.6477\\0.6452\\0.6466\\0.6479\\0.6492\\0.6428\\0.6483\\0.6489\\0.6475\\0.6454\\0.6493\\0.6462\\0.6470\\0.6360\\0.6369\\0.6337\\0.6402\\0.6370\\0.6333}&
\makecell{\textbf{\textcolor{red}{0.6105}}\\0.6076\\0.6054\\0.6065\\0.6079\\0.6086\\0.6031\\0.6092\\0.6095\\0.6075\\0.6057\\0.6096\\0.6058\\0.6070\\0.5925\\0.5947\\0.5937\\0.5973\\0.5963\\0.5940}&
\makecell{0.3739\\0.3725\\0.3732\\0.3712\\0.3727\\0.3727\\0.3704\\0.3726\\0.3719\\0.3721\\0.3765\\0.3760\\0.3757\\\textbf{\textcolor{blue}{0.3772}}\\0.3707\\0.3707\\0.3707\\0.3712\\0.3717\\0.3705}&
\makecell{0.3478\\0.3462\\0.3474\\0.3452\\0.3468\\0.3468\\0.3443\\0.3464\\0.3459\\0.3460\\0.3508\\0.3507\\0.3502\\\textbf{\textcolor{red}{0.3516}}\\0.3447\\0.3445\\0.3448\\0.3452\\0.3458\\0.3448}&
\makecell{0.6123\\0.6095\\0.6121\\0.6140\\0.6131\\0.6130\\0.6156\\0.6127\\0.5899\\0.5906\\0.6208\\\textbf{\textcolor{blue}{0.6209}}\\0.6182\\0.6181\\0.6138\\0.6107\\0.6108\\0.6120\\0.6107\\0.6116}&
\makecell{0.5868\\0.5841\\0.5869\\0.5884\\0.5874\\0.5876\\0.5913\\0.5864\\0.5688\\0.5694\\0.5958\\\textbf{\textcolor{red}{0.5960}}\\0.5941\\0.5939\\0.5889\\0.5859\\0.5862\\0.5869\\0.5865\\0.5868}&
\makecell{0.8744\\0.8719\\0.8734\\0.8700\\0.8755\\0.8730\\0.8672\\0.8835\\0.8850\\0.8794\\0.8910\\0.8922\\0.8953\\0.8961\\\textbf{\textcolor{blue}{0.9042}}\\0.8986\\0.8975\\0.9028\\0.8968\\0.8893}&
\makecell{0.8115\\0.8087\\0.8108\\0.8068\\0.8134\\0.8100\\0.8035\\0.8226\\0.8241\\0.8177\\0.8312\\0.8326\\0.8365\\0.8378\\\textbf{\textcolor{red}{0.8482}}\\0.8417\\0.8399\\0.8461\\0.8389\\0.8296}&
\makecell{0.7342\\0.7333\\0.7345\\0.7339\\0.7342\\0.7341\\0.7351\\0.7346\\0.7343\\0.7346\\0.7354\\0.7357\\0.7354\\\textbf{\textcolor{blue}{0.7362}}\\0.7354\\0.7344\\0.7343\\0.7358\\0.7355\\0.7348}&
\makecell{0.7052\\0.7041\\0.7057\\0.7050\\0.7053\\0.7051\\0.7068\\0.7058\\0.7055\\0.7058\\0.7070\\0.7074\\0.7072\\\textbf{\textcolor{red}{0.7083}}\\0.7070\\0.7056\\0.7055\\0.7075\\0.7071\\0.7061}

\\ \midrule

\makecell{FPN} &
\makecell{ResNet-50\\
          ResNet-101\\
          ResNeXt50\_32x4d\\
          ResNeXt101\_32x8d\\
          Res2Net-50\_26w\_4s\\
          Res2Net-101\_26w\_4s\\
          Vgg16\\
          Densenet-121\\
          Densenet-169\\
          Densenet-201\\
          SE-ResNet-50\\
          SE-ResNet-101\\
          SE-ResNeXt50\_32x4d\\
          SE-ResNeXt101\_32x4d\\
          RegNetx-002\\
          RegNetx-004\\
          RegNetx-006\\
          RegNety-002\\
          RegNety-004\\
          RegNety-006}&
\makecell{0.5505\\0.6313\\0.4737\\0.6329\\0.6341\\0.6338\\0.4931\\\textbf{\textcolor{blue}{0.6381}}\\0.6349\\0.6374\\0.6306\\0.6319\\0.6362\\0.4752\\0.5380\\0.4633\\0.3881\\0.6152\\0.6212\\0.3117}&
\makecell{0.5150\\0.5902\\0.4434\\0.5915\\0.5931\\0.5925\\0.4546\\\textbf{\textcolor{red}{0.5968}}\\0.5942\\0.5963\\0.5896\\0.5902\\0.5948\\0.4445\\0.5014\\0.4324\\0.3632\\0.5727\\0.5794\\0.2922}&
\makecell{0.2478\\0.2677\\0.2705\\0.2692\\\textbf{\textcolor{blue}{0.3177}}\\0.2765\\0.2478\\0.2907\\0.2722\\0.2715\\0.3176\\0.2961\\0.2478\\0.2478\\0.2478\\0.2478\\0.2478\\0.2712\\0.2478\\0.2478}&
\makecell{0.2458\\0.2611\\0.2636\\0.2624\\\textbf{\textcolor{red}{0.3013}}\\0.2675\\0.2458\\0.2792\\0.2652\\0.2644\\0.3012\\0.2845\\0.2458\\0.2458\\0.2458\\0.2458\\0.2458\\0.2643\\0.2458\\0.2458}&
\makecell{0.4995\\0.4995\\0.4995\\0.4995\\0.4995\\0.4995\\0.4995\\0.4995\\0.4995\\\textbf{\textcolor{blue}{0.4997}}\\0.4995\\0.4995\\0.4995\\0.4995\\0.4995\\0.4995\\0.4995\\0.4995\\0.4995\\0.4995}&
\makecell{0.4990\\0.4990\\0.4990\\0.4990\\0.4990\\0.4990\\0.4990\\0.4990\\0.4990\\\textbf{\textcolor{red}{0.4991}}\\0.4990\\0.4990\\0.4990\\0.4990\\0.4990\\0.4990\\0.4990\\0.4990\\0.4990\\0.4990}&
\makecell{0.7282\\0.5621\\0.5664\\0.6458\\0.8032\\0.6343\\0.4882\\0.8844\\\textbf{\textcolor{blue}{0.8935}}\\0.8928\\0.6524\\0.6433\\0.5705\\0.5720\\0.4882\\0.5713\\0.5711\\0.4882\\0.4882\\0.5650}&
\makecell{0.6873\\0.5403\\0.5453\\0.6146\\0.7515\\0.6016\\0.4776\\0.8231\\\textbf{\textcolor{red}{0.8342}}\\0.8337\\0.6226\\0.6116\\0.5504\\0.5525\\0.4776\\0.5517\\0.5514\\0.4776\\0.4776\\0.5436}&
\makecell{0.7314\\0.7303\\0.5507\\0.6310\\0.7293\\0.7313\\0.2355\\0.7331\\0.7324\\0.7324\\0.6921\\0.6922\\0.7325\\\textbf{\textcolor{blue}{0.7334}}\\0.3639\\0.6309\\0.6320\\0.6834\\0.6316\\0.5331}&
\makecell{0.7012\\0.6997\\0.5240\\0.6043\\0.6981\\0.7011\\0.2241\\0.7035\\0.7028\\0.7026\\0.6622\\0.6624\\0.7028\\\textbf{\textcolor{red}{0.7038}}\\0.3381\\0.6038\\0.6055\\0.6525\\0.6049\\0.5104}

\\ \bottomrule
\end{tabular}
}
\label{tab:tests1}
\end{table*}

\section{Architectures and Datasets}
\label{architectures}

In order to evaluate different networks in the problem of Covid-19 \gls*{ct} segmentation, we use networks based on encoder-decoder structure. We evaluated twenty encoders: ResNet-50 and ResNet-101~\cite{He2016}; ResNeXt50 and ResNeXt101~\cite{Xie2017}; Res2Net50 and Res2Net101~\cite{Gao2021}; VGG16~\cite{simonyan2015deep}; DenseNet121, DenseNet169, and DenseNet201~\cite{Huang2017}; \gls*{se}-ResNet50, \gls*{se}-ResNet101, \gls*{se}-ResNeXt50, and \gls*{se}-ResNeXt101~\cite{Hu2018}; RegNetx-002, RegNetx-004, RegNetx-006, RegNety-002, RegNety-004, and RegNety-006~\cite{Radosavovic2020}; and six decoders: U-Net~\cite{Ronneberger2015}, U-Net++~\cite{Zhou2018}, \gls*{fpn}~\cite{Lin2017}, \gls*{pspnet}~\cite{Zhao2017}, LinkNet~\cite{Chaurasia2017}, and \gls*{manet}~\cite{Fan2020manet}. Each encoder is combined with all decoders, resulting in a total of 120 networks evaluated. 

The networks were evaluated through the F-score, see equation~\ref{eq:dice_loss}, and \gls*{iou}, see equation~\ref{eq:jaccard_loss}, as metrics and trained with the loss function presented in equation~\ref{eq:weighted_mean}:

\begin{equation}
     Loss = 1 - \frac{w_{1} * F-score + w_{2} * IoU}{2}
\label{eq:weighted_mean}
\end{equation}

where:

\begin{equation}
    F-score = \frac{TruePositive}{TruePositive + \frac{FalsePositive + FalseNegative}{2}}
\label{eq:dice_loss}
\end{equation}

and

\begin{equation}
     IoU = \frac{intersection}{union}
\label{eq:jaccard_loss}
\end{equation}

The weights $w_{1}$ and $w_{2}$ were defined as 1. The benchmark code for running these same experiments is publicly available on\footnote{https://github.com/VRI-UFPR/SparkInTheDarkLars2021}. 


\begin{table*}[!ht]
\centering
\caption{The average value of five-folds cross-validation strategy for test set evaluated with the last train weight. The best F-score values are highlighted in blue and the best IoU values are highlighted in red (Continuation of table~\ref{tab:tests1}).} 
\resizebox{0.9\textwidth}{!}{%
\begin{tabular}{cccccccccccc}

\toprule
\multicolumn{1}{c}{\textbf{Decoder}} &
\multicolumn{1}{c}{\textbf{Encoder}} &
\multicolumn{2}{c}{\textbf{CC-CCII}} &
\multicolumn{2}{c}{\textbf{MedSeg}} &
\multicolumn{2}{c}{\textbf{MosMed}} &
\multicolumn{2}{c}{\textbf{Ricord1a}} &
\multicolumn{2}{c}{\textbf{Zenodo}} 
\\ \midrule

\makecell{} &
\makecell{} &
\makecell{F-score} &
\makecell{IoU} & 
\makecell{F-score} &
\makecell{IoU} & 
\makecell{F-score} &
\makecell{IoU} & 
\makecell{F-score} &
\makecell{IoU} & 
\makecell{F-score} &
\makecell{IoU}
\\ \midrule

\makecell{PSPNet} &
\makecell{ResNet-50\\
          ResNet-101\\
          ResNeXt50\_32x4d\\
          ResNeXt101\_32x8d\\
          Res2Net-50\_26w\_4s\\
          Res2Net-101\_26w\_4s\\
          Vgg16\\
          Densenet-121\\
          Densenet-169\\
          Densenet-201\\
          SE-ResNet-50\\
          SE-ResNet-101\\
          SE-ResNeXt50\_32x4d\\
          SE-ResNeXt101\_32x4d\\
          RegNetx-002\\
          RegNetx-004\\
          RegNetx-006\\
          RegNety-002\\
          RegNety-004\\
          RegNety-006}&
\makecell{0.4822\\0.4817\\0.5366\\0.4823\\0.5044\\0.4822\\0.4816\\0.4992\\0.4995\\0.4933\\\textbf{\textcolor{blue}{0.5420}}\\0.5245\\0.4832\\0.4826\\0.4756\\0.4783\\0.4800\\0.4764\\0.4799\\0.4808}&
\makecell{0.4698\\0.4692\\0.5122\\0.4701\\0.4872\\0.4700\\0.4687\\0.4831\\0.4832\\0.4788\\\textbf{\textcolor{red}{0.5153}}\\0.5020\\0.4708\\0.4702\\0.4589\\0.4628\\0.4659\\0.4596\\0.4657\\0.4668}&
\makecell{0.3598\\0.2942\\0.3161\\0.3616\\0.3119\\0.3337\\0.3394\\0.2843\\\textbf{\textcolor{blue}{0.3621}}\\0.3162\\0.3620\\0.3443\\0.3338\\0.3549\\0.2478\\0.2478\\0.2478\\0.2478\\0.2478\\0.2478}&
\makecell{0.3336\\0.2814\\0.2996\\0.3354\\0.2958\\0.3129\\0.3159\\0.2743\\\textbf{\textcolor{red}{0.3358}}\\0.2997\\\textbf{\textcolor{red}{0.3358}}\\0.3215\\0.3130\\0.3305\\0.2458\\0.2458\\0.2458\\0.2458\\0.2458\\0.2458}&
\makecell{0.5201\\0.5950\\0.5846\\0.5631\\0.5832\\0.5620\\0.5803\\0.5632\\0.5520\\0.5564\\0.5831\\0.5550\\0.5428\\\textbf{\textcolor{blue}{0.6080}}\\0.4995\\0.4995\\0.4995\\0.4995\\0.4995\\0.4995}&
\makecell{0.5145\\0.5704\\0.5634\\0.5471\\0.5623\\0.5461\\0.5592\\0.5470\\0.5375\\0.5414\\0.5620\\0.5404\\0.5318\\\textbf{\textcolor{red}{0.5815}}\\0.4990\\0.4990\\0.4990\\0.4990\\0.4990\\0.4990}&
\makecell{0.8881\\0.8858\\0.8917\\0.8901\\0.8910\\0.8908\\0.8793\\0.8940\\0.8930\\0.8920\\0.8925\\0.8913\\0.8941\\\textbf{\textcolor{blue}{0.8957}}\\0.8418\\0.8601\\0.8747\\0.8451\\0.8764\\0.8788}&
\makecell{0.8276\\0.8247\\0.8322\\0.8303\\0.8313\\0.8310\\0.8171\\0.8350\\0.8337\\0.8324\\0.8329\\0.8315\\0.8355\\\textbf{\textcolor{red}{0.8369}}\\0.7731\\0.7944\\0.8118\\0.7769\\0.8136\\0.8168}&
\makecell{0.6192\\0.6386\\0.6614\\0.6406\\0.6412\\0.6612\\\textbf{\textcolor{blue}{0.7001}}\\0.6200\\0.6198\\0.6198\\0.6411\\0.6621\\0.6623\\0.6197\\0.6470\\0.5197\\0.6364\\0.6288\\0.6170\\0.6172}&
\makecell{0.5993\\0.6156\\0.6364\\0.6183\\0.6190\\0.6360\\\textbf{\textcolor{red}{0.6690}}\\0.6004\\0.6002\\0.6001\\0.6190\\0.6372\\0.6378\\0.6003\\0.6162\\0.4970\\0.6122\\0.6013\\0.5959\\0.5963}

\\ \midrule

\makecell{LinkNet} &
\makecell{ResNet-50\\
          ResNet-101\\
          ResNeXt50\_32x4d\\
          ResNeXt101\_32x8d\\
          Res2Net-50\_26w\_4s\\
          Res2Net-101\_26w\_4s\\
          Vgg16\\
          Densenet-121\\
          Densenet-169\\
          Densenet-201\\
          SE-ResNet-50\\
          SE-ResNet-101\\
          SE-ResNeXt50\_32x4d\\
          SE-ResNeXt101\_32x4d\\
          RegNetx-002\\
          RegNetx-004\\
          RegNetx-006\\
          RegNety-002\\
          RegNety-004\\
          RegNety-006}&
\makecell{0.6432\\0.6438\\0.6281\\0.6455\\0.6448\\0.6429\\0.6414\\0.6142\\0.6275\\0.6353\\0.6461\\0.6455\\\textbf{\textcolor{blue}{0.6467}}\\0.6449\\0.6135\\0.6185\\0.6215\\0.6135\\0.6316\\0.6286}&
\makecell{0.6023\\0.6030\\0.5878\\0.6049\\0.6044\\0.6023\\0.6024\\0.5707\\0.5862\\0.5941\\0.6062\\0.6056\\\textbf{\textcolor{red}{0.6064}}\\0.6054\\0.5704\\0.5766\\0.5810\\0.5705\\0.5906\\0.5884}&
\makecell{0.3561\\0.3632\\0.3688\\0.3696\\0.3666\\0.3695\\0.3715\\0.3708\\0.3629\\0.3693\\0.3743\\0.3753\\0.3757\\\textbf{\textcolor{blue}{0.3761}}\\0.3601\\0.3671\\0.3673\\0.3659\\0.3675\\0.3687}&
\makecell{0.3288\\0.3328\\0.3412\\0.3432\\0.3375\\0.3431\\0.3453\\0.3445\\0.3314\\0.3430\\0.3482\\0.3496\\0.3500\\\textbf{\textcolor{red}{0.3505}}\\0.3333\\0.3409\\0.3415\\0.3396\\0.3411\\0.3427}&
\makecell{0.6111\\0.6072\\0.6125\\0.6103\\0.6096\\0.6085\\0.6136\\0.6076\\0.6076\\0.6056\\0.6207\\0.6211\\0.6212\\\textbf{\textcolor{blue}{0.6213}}\\0.6038\\0.6085\\0.6088\\0.6063\\0.6070\\0.6127}&
\makecell{0.5857\\0.5817\\0.5868\\0.5843\\0.5838\\0.5824\\0.5888\\0.5816\\0.5820\\0.5800\\0.5959\\0.5963\\0.5964\\\textbf{\textcolor{red}{0.5965}}\\0.5793\\0.5836\\0.5836\\0.5808\\0.5811\\0.5869}&
\makecell{0.8612\\0.8563\\0.8643\\0.8643\\0.8679\\0.8716\\0.8647\\0.8755\\0.8730\\0.8702\\0.8893\\0.8895\\0.8934\\0.8944\\0.8913\\0.8935\\0.8895\\0.8911\\\textbf{\textcolor{blue}{0.8946}}\\0.8934}&
\makecell{0.7960\\0.7903\\0.7998\\0.7997\\0.8040\\0.8078\\0.8002\\0.8129\\0.8101\\0.8065\\0.8288\\0.8292\\0.8339\\0.8352\\0.8318\\0.8346\\0.8296\\0.8311\\\textbf{\textcolor{red}{0.8357}}\\0.8346}&
\makecell{0.7323\\0.7315\\0.7339\\0.7322\\0.7323\\0.7324\\0.7353\\0.7331\\0.7329\\0.7332\\0.7096\\0.7354\\\textbf{\textcolor{blue}{0.7361}}\\\textbf{\textcolor{blue}{0.7361}}\\0.7250\\0.7328\\0.7303\\0.7314\\0.7335\\0.7343}&
\makecell{0.7031\\0.7018\\0.7051\\0.7029\\0.7030\\0.7031\\0.7070\\0.7039\\0.7038\\0.7040\\0.6779\\0.7070\\\textbf{\textcolor{red}{0.7081}}\\\textbf{\textcolor{red}{0.7081}}\\0.6926\\0.7032\\0.6998\\0.7011\\0.7043\\0.7053}

\\ \midrule

\makecell{MA-Net} &
\makecell{ResNet-50\\
          ResNet-101\\
          ResNeXt50\_32x4d\\
          ResNeXt101\_32x8d\\
          Res2Net-50\_26w\_4s\\
          Res2Net-101\_26w\_4s\\
          Vgg16\\
          Densenet-121\\
          Densenet-169\\
          Densenet-201\\
          SE-ResNet-50\\
          SE-ResNet-101\\
          SE-ResNeXt50\_32x4d\\
          SE-ResNeXt101\_32x4d\\
          RegNetx-002\\
          RegNetx-004\\
          RegNetx-006\\
          RegNety-002\\
          RegNety-004\\
          RegNety-006}&
\makecell{0.6468\\0.6448\\0.6461\\0.6462\\0.6468\\0.6475\\0.6452\\0.6470\\0.6455\\0.6470\\\textbf{\textcolor{blue}{0.6484}}\\0.6479\\0.6431\\0.6432\\0.6402\\0.6392\\0.6372\\0.6440\\0.6395\\0.6351}&
\makecell{0.6036\\0.6023\\0.6047\\0.6052\\0.6051\\0.6062\\0.6047\\0.6055\\0.6039\\0.6057\\\textbf{\textcolor{red}{0.6076}}\\0.6075\\0.6026\\0.6031\\0.5983\\0.5987\\0.5969\\0.6020\\0.5988\\0.5949}&
\makecell{0.3682\\0.3692\\0.3709\\0.3712\\0.3701\\0.3669\\0.3737\\0.3645\\0.3659\\0.3647\\\textbf{\textcolor{blue}{0.3759}}\\0.3758\\0.3751\\0.3755\\0.3725\\0.3714\\0.3721\\0.3715\\0.3713\\0.3710}&
\makecell{0.3422\\0.3430\\0.3449\\0.3446\\0.3438\\0.3406\\0.3474\\0.3381\\0.3389\\0.3383\\\textbf{\textcolor{red}{0.3502}}\\0.3501\\0.3497\\0.3498\\0.3467\\0.3458\\0.3461\\0.3457\\0.3453\\0.3454}&
\makecell{0.6132\\0.6109\\0.6133\\0.6086\\0.6122\\0.6109\\0.6169\\0.6099\\0.5797\\0.6084\\\textbf{\textcolor{blue}{0.6212}}\\0.6206\\0.6172\\0.6197\\0.6130\\0.6089\\0.6118\\0.6105\\0.6109\\0.6101}&
\makecell{0.5880\\0.5856\\0.5875\\0.5831\\0.5866\\0.5852\\0.5923\\0.5840\\0.5599\\0.5823\\\textbf{\textcolor{red}{0.5960}}\\0.5956\\0.5928\\0.5948\\0.5880\\0.5844\\0.5873\\0.5859\\0.5862\\0.5852}&
\makecell{0.8691\\0.8660\\0.8733\\0.8634\\0.8689\\0.8741\\0.8789\\0.8565\\0.8597\\0.8615\\0.8819\\0.8873\\\textbf{\textcolor{blue}{0.8918}}\\0.8917\\0.8888\\0.8874\\0.8817\\0.8825\\0.8835\\0.8851}&
\makecell{0.8055\\0.8017\\0.8104\\0.7982\\0.8053\\0.8112\\0.8171\\0.7907\\0.7944\\0.7965\\0.8208\\0.8265\\0.8321\\\textbf{\textcolor{red}{0.8323}}\\0.8288\\0.8269\\0.8204\\0.8210\\0.8223\\0.8241}&
\makecell{0.7310\\0.7306\\0.7315\\0.7313\\0.7312\\0.7308\\0.7351\\0.7304\\0.7296\\0.7301\\0.7331\\0.7333\\0.7335\\\textbf{\textcolor{blue}{0.7352}}\\0.7345\\0.7340\\0.7332\\0.7336\\0.7343\\0.7330}&
\makecell{0.7008\\0.7002\\0.7016\\0.7013\\0.7011\\0.7004\\\textbf{\textcolor{red}{0.7067}}\\0.7002\\0.6990\\0.6997\\0.7038\\0.7041\\0.7043\\\textbf{\textcolor{red}{0.7067}}\\0.7053\\0.7048\\0.7038\\0.7043\\0.7054\\0.7034}
          
\\ \bottomrule
\end{tabular}
}
\label{tab:tests2}
\end{table*}

The models were trained and evaluated across five different \glspl*{ct} datasets: MedSeg~\cite{medseg}, Zenodo~\cite{zenodo}, CC-CCII~\cite{Zhang2020}, MosMed~\cite{Morozov2020}, and Ricord1a~\cite{ricord1a}. The MedSeg was one of the first datasets proposed in the literature, being composed of 929 images and labels for four classes, with the following pixel proportion: Background (0.98563), \gls*{ggo} (0.01072), Consolidation (0.00351), and Pleural Effusion (0.0001). The Zenodo dataset, an evolution of MedSeg, is composed of 3,520 images and also has labels for four classes, with the following pixel proportion: Background (0.89893), Left Lung (0.04331), Right Lung (0.04923), and Infections (0.00852). The MosMed dataset is composed of 2,049 images, with labels for two classes, with the following pixel proportion: Background (0.99810) and \gls*{ggo}-Consolidation (0.00189). Ricord dataset is divided into three sets: 1a, 1b, and 1c. The set 1a is the only one with segmentation masks and used in our work. The Ricord1a is the largest dataset, being composed of 9,166 images and has labels for two classes, with the following pixel proportion: Background  (0.95295) and Infections (0.04704). We also used a sub-set of CC-CCII with segmentation masks. This sub-set is composed of 750 images and has labels for four classes, with the following pixel proportion: Background (0.87152), Lung Field (0.11691), \gls*{ggo} (0.00802), and Consolidation (0.00353).

\section{Experiments}

First, all datasets were divided into training and test sets, with the training set being composed of 80\% of the images and the test set 20\% of the images. Then, all networks were trained for 50 epochs in the training set. Transfer learning strategy was used, and the network weights were initialized with the ImageNet~\cite{Deng2009} weights. We used a batch size of 8, with the learning rate starting at 0.001 and being divided by 10 every 10 epochs. In order to evaluate only the networks, no data augmentation was applied. The input images were resized to 256x256. The scaling factor was based on the largest dimension of the image to avoid distortion, and the smallest one was padded with zeros accordingly. Each encoder-decoder combination was trained in five datasets, with each dataset being validated through a five-fold cross-validation strategy, totaling 3.000 experiments. In Fig.~\ref{fig:training} the curves are color-coded by decoder (U-Net, U-Net++, \gls*{fpn}, \gls*{pspnet}, LinkNet, and \gls*{manet}) and represent the average F-score of the five-folds of all encoders (ResNet-50-fold0, ..., ResNet-50-fold4, ResNet-101-fold0, ..., ResNet-101-fold4, etc.) using that decoder.


The U-Net++, U-Net, LinkNet, and \gls*{manet} achieved similar results, with the U-Net++ reaching the best results in most of the datasets. The \gls*{fpn} and \gls*{pspnet} achieved the worst results. However, the \gls*{pspnet} reached close results with the U-Net++ in the Ricord1a dataset. Also, all networks achieved great generalization, with the validation curve being very close to the training curve in most of the datasets. In the CC-CCII dataset, the networks did not achieve such generalization, and the validation curve is far from the train in most networks, with the \gls*{pspnet} being the only network that had good generalization in this dataset.

Also, the \gls*{fpn} was not capable of learning the \gls*{ggo}-Consolidation label in the MosMed dataset. As presented in Fig. 1, both train and validation curves are straight lines with an F-score of 50\%. The \gls*{fpn} achieved an F-score of 99\% for background and 0\% for lesion class. This behavior is due to the critical class imbalance present in this dataset. The number of classes in this dataset (two classes) also contributed to this result. In the Ricord1a dataset, the \gls*{fpn} achieved results very far from the other networks.

Table~\ref{tab:tests1} and Table~\ref{tab:tests2} presents the test results obtained using the average value of the five-fold cross-validation strategy performed with the last train weight. The best F-score values are highlighted in blue, and the best IoU values are highlighted in red. The U-Net, U-Net++, and \gls*{manet} presented more stable results, with very close results between the many encoders evaluated. However, the \gls*{fpn}, \gls*{pspnet}, and LinkNet showed to be more sensitive to the encoder change, with a higher variation on F-score and IoU. The \gls*{fpn} and \gls*{pspnet} achieved the worst results among the decoders, while the other decoders achieved very close results. 

The MedSeg was the most challenging dataset, with all networks achieving the lowest results in this dataset. However, when compared with other datasets of Covid-19 segmentation, the MedSeg dataset presents some issues like coarse segmentation masks with the segmentation bypassing the lung region and trash markings like circles and arrows pointing to the lesion region. These issues difficult the learning process. 

In order to perform a statistical analysis of the trained models, we applied the Friedman test with the null hypothesis that the models have the same distribution. We used the individual F-score value of each image in the test set to compose the distribution of each trained model. Both encoders and decoders were compared. First, for each decoder, the Friedman test was applied to verify if there is a statistical difference between the encoders trained with that decoder. Than, the Friedman test was applied again to verify if there is a statistical difference between decoders. 

In both comparisons, the resulting $p$-value of the Friedman test was too small to represent a double variable and end up zero, so we refuse the hypothesis and affirm that there is a statistical difference between the encoders and decoders. So, besides the average F-score and IoU of each model being close, changing the encoders linked to the decoder generates a different distribution in the test images..





\section{Conclusion}

As expected, there is no definitive answer for which encoder-decoder combination is the best to be applied in the approached problem. This study provides as the main contribution a robust guideline for future works as encoder-decoder selection, setting of variables like the number of network parameters, and training time. In terms of decoders, the U-Net and U-Net++, widely applied in the literature, achieved impressive results. However, the LinkNet and the \gls*{manet} achieved very close results to them. Also, the \gls*{fpn} and \gls*{pspnet} presented the worst results, appearing to be unsuitable for this problem. In terms of encoders, the performance varied depending on decoder combination and dataset applied. In general, they achieved close results, with none of them standing out.

Also, this analysis revealed some weak points to be approached in future works in terms of datasets. The major problem  is the critical class imbalance within the datasets, which affects the learning process of deep neural networks. Future works could investigate data augmentation and other balancing techniques to mitigate this problem.

\section*{ACKNOWLEDGMENT}
The authors would like to thank the Coordination for the Improvement of Higher Education Personnel  (CAPES) for the PhD scholarship. We gratefully acknowledge the founders of the publicly available datasets, the support of NVIDIA Corporation with the donation of the GPUs used for this research and the C3SL-UFPR  group for the computational cluster infrastructure.





\bibliographystyle{IEEEtran}
\bibliography{references}

\end{document}